\documentclass[oldversion]{aa}
\usepackage{txfonts}
\usepackage{graphicx}
\usepackage{natbib}
\usepackage{supertabular}

\bibpunct{(}{)}{;}{a}{}{,}

\begin{document}
\title{Microlensing variability in time-delay quasars}
\author{Danuta Paraficz \inst{1,2} \and Jens Hjorth \inst{1} \and Ingunn Burud \inst{3} \and P\'all Jakobsson \inst{1}
\and \'Ard{\'\i}s El{\'\i}asd\'ottir \inst{1}}
\offprints{D. Paraficz}

\institute{Dark Cosmology Centre, Niels Bohr Institute, University of Copenhagen,
		Juliane Maries Vej 30, DK-2100 Copenhagen, Denmark
              \email{danutas@astro.ku.dk}
\and
		Nordic Optical Telescope (NOT), Apartado 474, 38700 Santa Cruz de La Palma, Canary Islands, Spain     
\and
                Norwegian Meteorological Institute, P.O. Box 43, Blindern, N-031 3 Oslo, Norway
}     

\date{Received March 26, 2006}
\abstract
{We have searched for microlensing variability in the light curves of five gravitationally lensed quasars
with well-determined time delays: SBS~1520+530, FBQ~0951+2635, RX~J0911+0551, 
B1600+434 and HE~2149$-$2745.
By comparing the  light curve of the leading image with a suitably time offset light curve of a
trailing image we find that two (SBS~1520+530 and FBQ~0951+2635) out of the five quasars have 
significant long-term ($\sim$years) and short-term ($\sim$100 days) brightness variations 
that may be attributed to microlensing.
The short-term variations may be due to nanolenses 
$10^{-4}$--$10^{-3}$ M$_\odot$,
relativistic hot or cold spots in the quasar accretion disks,
or coherent microlensing at large optical depth.}


\maketitle

\section{Introduction}

The effect of a background source being gravitationally lensed by foreground compact objects is known 
as microlensing. \citet{CHR79} predicted that in lensed quasar systems the light path should be 
affected by stars in the lensing galaxy. Moving compact objects in the lensing galaxy can cause
spectral changes, brightness variability and, in the case of multiple images, flux ratio 
anomalies in the lensed quasar. When the foreground galaxy causes multiple images of the quasar, 
photometric monitoring can be used to isolate intrinsic quasar variability from microlensing 
variability by comparison of the separate light curves. Likewise, spectral differences between quasar 
images, caused by differential magnification across the quasar, can be directly detected in multiply 
imaged quasars. Microlensing offers the opportunity to study the nature of matter in foreground 
galaxies and the spatial structure of the lensed quasar at very high angular resolution \citep{KO04}. 
The first quasar microlensing events were discovered 17 years ago by \citet{IR89} and \citet{Schil90}.

In this Letter we analyze the light curves of f\mbox{}ive lensed quasars for 
which a time delay  has previously been measured in dedicated monitoring 
campaigns \citep{zzz00,BC02,zzz02,HB02,zz05}.
Four of them were observed at the Nordic Optical Telescope (NOT) between 
1998--2002 and one (HE~2149$-$2745) was observed at the Danish 1.5-m telescope. 
 We here examine the light curves for microlensing variability and
at the same time make all data points available online.

\begin{table}[!t]
\label{table:1}
\caption{Properties of the lens systems. The time delays were obtained by: 1. \citet{zzz02}, 2. \citet{zz05}, 3. \citet{BC02}, 4. \citet{HB02}, 5. \citet{zzz00}.}
\centering
\begin{tabular}{l l l l l l}
\hline\hline
Name& $z_l$&$z_s$&$\Delta t$ (days)&Separation & Ref.\\
\hline
SBS 1520+530&0.72&1.86&$130\pm 2$ &1.57\arcsec\ & 1 \\
FBQ 0951+2635& $0.24$&$1.25$&$16\pm 2$ &1.10\arcsec\ & 2 \\
HE~2149$-$2745&0.495  & 2.03  &$103\pm12$ &1.70\arcsec\ & 3 \\
RX~J0911+0551& 0.77  &  2.80 &$146 \pm 4$ &3.10\arcsec\ & 4 \\
B1600+434&     0.41  &  1.59 &$51\pm2$ &1.38\arcsec\ & 5 \\
\hline
\end{tabular}
\end{table}
\section{Analysis}
 
The images of a lensed quasar may vary due to intrinsic quasar brightness changes and/or
microlensing.
Microlensing affects the light paths of each image
differently (in the simplest case only one path is affected) whereas the
intrinsic variations show up in all the images but at different times due to
the time delay.
Therefore, one can isolate the microlensing signal by calculating the difference between
two light curves, suitably shifted in time to correct for the time delay.

Observationally, quasar light curves suffer from sampling effects leading to a need for
interpolation. Due to their small separations and the presence of the lensing galaxy they may also potentially be affected by 
systematic errors in the photometry. To eliminate artifacts arising from such effects 
we do not interpolate data points with a gap bigger then a time delay of a system and we require that the light curve difference be
(1) uncorrelated with the quasar variations and
(2) independent of which image light curve we interpolate.
These are conservative criteria for detection of a microlensing signal in 
the light curves which ensure that the signal is reliable and robust. 
If the criteria are violated the variability may be the result 
of observational, systematic or interpolation errors.

\section{Results}

Were we apply the analysis to the f\mbox{}ive quasar light curves to search for microlensing events.

\subsection{SBS 1520+530}
\begin{figure}
\resizebox{\hsize}{!}{\includegraphics{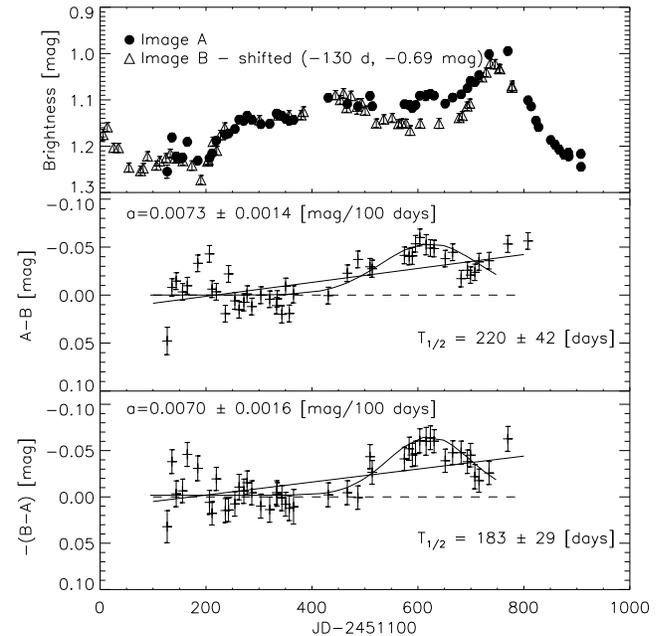}}
\caption{R-band light curves of SBS 1520+530.
\textbf{Top:} Time-delay shifted light curves, with the B image offset by
$-0.69$ mag and $-130$ days.  \textbf{Middle:} Difference between linearly interpolated A image
and B image. \textbf{Bottom:}
Difference between linearly interpolated B image and A image. 
The middle and bottom panels also present   linear fits (where $a$ is the slope)  and Gaussian fits 
(where \textbf{$T_{1/2}$} is the FWHM) to the data.}
\label{sbs}
\end{figure}

 The doubly imaged quasar SBS 1520+530 (see Table \ref{table:1}) was monitored between February 1999 and May 2001 at the NOT.
In Fig.~\ref{sbs} we show the light curves of the two images, A and B, where B has been shifted in both  
time ($-130$ days) and brightness ($-0.69$ mag).
Also shown are the magnitude residuals, $\Delta m$,
obtained by linear interpolation of one of the images.
 The middle panel shows the difference between a  linearly interpolated  A light curve and the B data points  ($\Delta m_{\mathrm{AB}}$). 
The lower panel shows the difference between a linearly interpolated  B light curve and the A data points ($\Delta m_{\mathrm{BA}}$).
Any deviation from a constant difference lightcurve (dashed horizontal
line) can be attributed to microclensing.
In both residual plots one notices a constant magnitude increase and an approximately 200-day-wide magnitude bump. 
In the following we characterize such residuals phenomenologically using 
linear or Gaussian f\mbox{}its.

A linear f\mbox{}it shows that the slopes of the plots are 0.0073$\pm$0.0014 mag/100 days
for $\Delta m_{\mathrm{AB}}$ and 0.0070$\pm$0.0016 mag/100 days for $\Delta m_{\mathrm{BA}}$.
 The agreement between these values indicates that
the mean difference between the images increases
 by about 0.007 mag per 100 days. 
This is consistent with the observations of \citet{G05} who found an increase in the
 magnitude difference between the quasar images of 0.14 $\pm$ 0.03 mag over 1500 days  (0.009 $\pm$ 0.002 mag/100 days).
We interpret this signal as evidence for microlensing.

A Gauss function  with baseline zero, f\mbox{}itted to the bump for $\Delta m_{\mathrm{AB}}$ gives a FWHM of $220\pm 42$ days and a peak  
magnitude variation of $0.053\pm 0.008$ mag. 
For $\Delta m_{\mathrm{BA}}$ the FWHM is $183\pm 29$ days and the peak magnitude variation is $0.063\pm 0.009$ mag. 
 These values are consistent within the errors.

 Microlensing in SBS~1520+530 was already detected spectroscopically by \citet{zzz02} who
analyzed continuum normalized spectra and showed that the equivalent widths of emission lines in A were larger than in B.

\subsection{FBQ 0951+2635}
The doubly imaged quasar FBQ 0951+2635 (see Table  \ref{table:1}) was observed between March 1999 and June 2001 at the NOT.
Spectroscopic indications of possible microlensing in the system were found by \citet{Sch1} and \citet{zz05}.

\begin{figure}
\resizebox{\hsize}{!}{\includegraphics{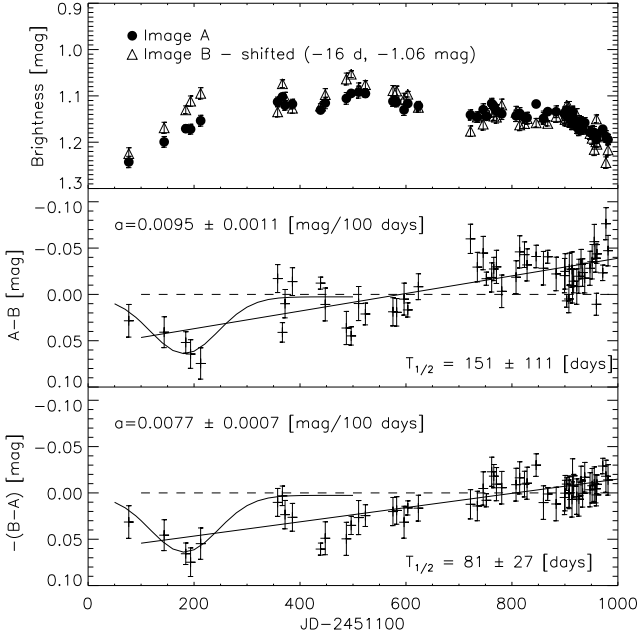}}
\caption{R-band light curves of FBQ 0951+2635. \textbf{Top:} Time-delay shifted light curves, with the B image 
 offset by $-1.06$ mag and $-16$ days.  \textbf{Middle:} Difference between linearly interpolated A image
and B image. \textbf{Bottom:} Difference between linearly interpolated B image and A image.
The middle and bottom panels also present   linear f\mbox{}its (where $a$ is the slope)  and Gaussian f\mbox{}its 
(where \textbf{$T_{1/2}$} is the FWHM) to the data.}
\label{fbq}

\end{figure}

We repeat the microlensing-extraction procedure described above for  FBQ 0951+2635.
 In Fig.~\ref{fbq} we show the light curves of the two images, A and B, the latter
shifted in time by $-16$  days and in brightness by $-1.06$ mag.
The middle and lower panels show $\Delta m_{\mathrm{AB}}$ and $\Delta m_{\mathrm{BA}}$.
 In both plots one notices a constant magnitude increase and a bump at the beginning
of the observations. 

Linear f\mbox{}its to the data yield slopes of 0.0095$\pm$0.0011 
  and 0.0077$\pm$0.0007 mag/100 days for $\Delta m_{\mathrm{AB}}$ and $\Delta m_{\mathrm{BA}}$ respectively.
 A Gauss function with baseline zero, f\mbox{}itted to the bump  for
$\Delta m_{\mathrm{AB}}$ gives a FWHM of 151 $\pm$ 111 days and a peak  magnitude variation of $0.048\pm 0.009$ mag.
For $\Delta m_{\mathrm{BA}}$ gives FWHM of 81 $\pm$  27 days and a peak  magnitude variation of $0.061\pm 0.016$ mag.. 
These values are consistent within the errors.

\subsection{Quasars where no microlensing is confirmed}
The three quasars where no microlensing variability was reliably detected are 
the double quasar HE~2149$-$2745 observed at the Danish 1.5-m telescope, 
ESO-La Silla (October 1998 -- December 2000), the quadruple quasar 
RX~J0911+0551 observed at the NOT (March 1997 -- April 2001), and the doubly 
imaged quasar B1600+434 observed at the NOT (April 1998 -- November 1999)
(see Table  \ref{table:1}).
\begin{figure}
\resizebox{\hsize}{!}{\includegraphics{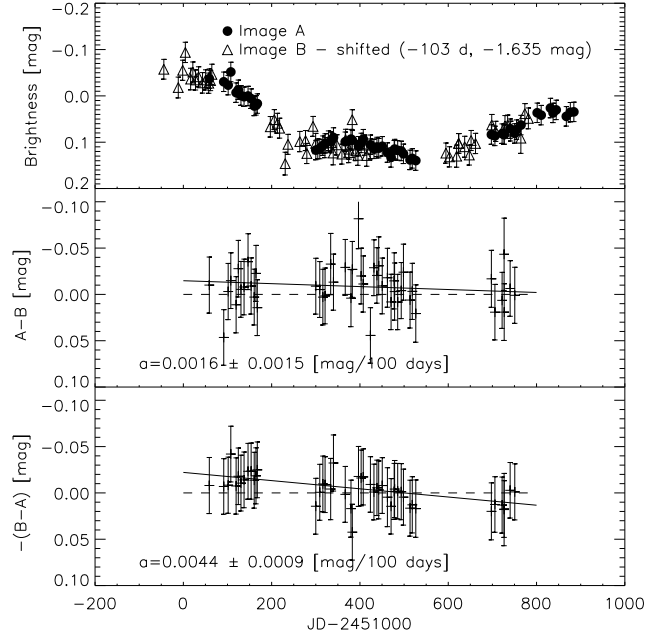}}

\caption{V-band light curves of HE~2149$-$2745 and the shifted light curve differences. 
The middle and bottom  panels include linear fits with $a$ being the slope of the linear fit.}
\label{hev}
\end{figure}

\begin{figure}
\resizebox{\hsize}{!}{\includegraphics{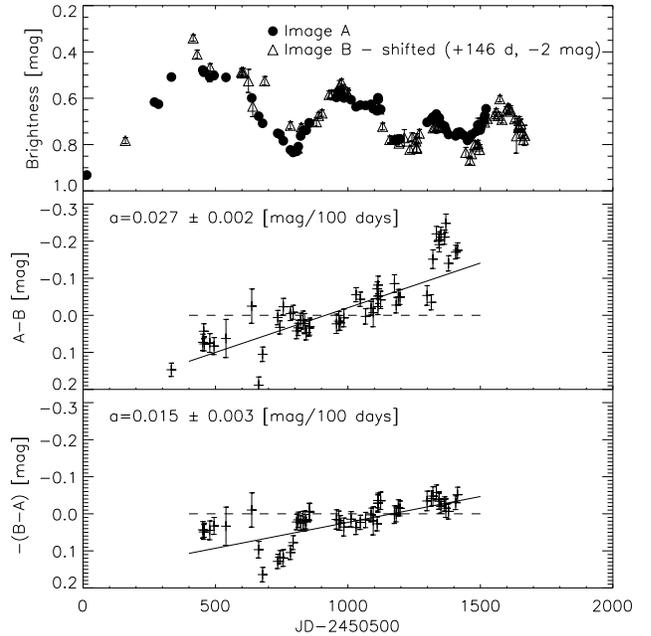}}
\caption{I-band light curves of RX~J0911+055 and the shifted light curve differences. 
The middle and bottom panels include linear fits with $a$ being the slope of the linear fit.}
\label{r}
\end{figure}

\begin{figure}
\resizebox{\hsize}{!}{\includegraphics{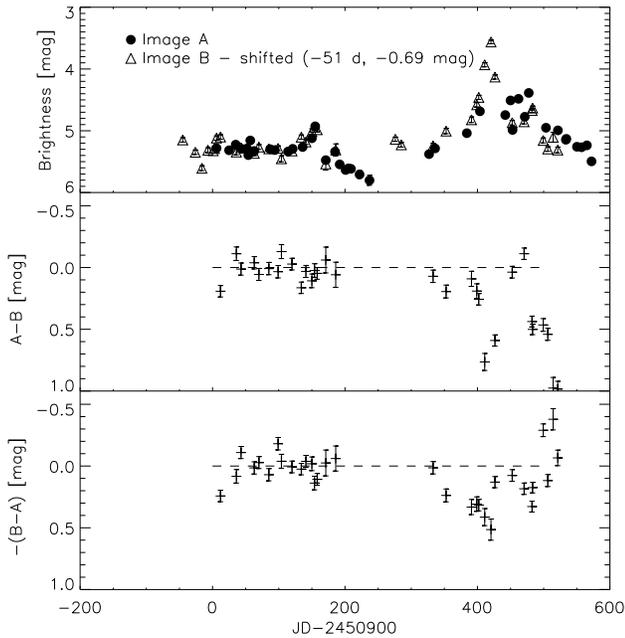}}
\caption{I-band light curves of B1600+434 and the shifted light curve differences. }
\label{b}
\end{figure}

Plots similar to Figs.~1 and 2 are shown in Figs.~3--5.
For HE~2149$-$2745 (Fig.~\ref{hev}) and RX~J0911+0551 (Fig.~\ref{r})   
we see some long term variations in the light curve difference but the analysis  
shows that the slopes differ depending on which of the two light curves are interpolated.
In B1600+434 (Fig.~\ref{b}) one sees clear magnitude variations in the light curve differences but they
do not satisfy our second condition for microlensing. By 
comparison with the top plot we see that they are correlated with the quasar variations 
 (the magnitude bump appears at $\sim 420$ days on all plots).

Thus, while we detect intriguing light curve differences and cannot rule out 
microlensing variability in these systems, this indicates that the detected 
magnitude changes may also be attributed to photometric or interpolation errors.

\section{Discussion}

 We have demonstrated the existence
 of a short-term microlensing variability in the photometric data sets of the two quasar systems SBS~1520+530 and FBQ~0951+2635. 
Interestingly, microlensing has previously been detected spectroscopically  and as flux anomalies
 in both systems  \citep{zzz02,faure02,Sch1,zz05}.  We note that the time scales of  the short-term variability detected in the quasars 
 (50--200 days) is similar to the '90 day' events in QSO Q0957+561A,B \citep[see][]{SCH96}.

In principle one can calculate the masses of the lenses responsible for the microlensing
from the time scales of brightness variations (FWHM) and the transverse velocities of the compact objects.
 An approximate value of the lens mass is given by 
\begin{equation}
M\approx\left(\frac{\rm{time}\times  V_{\rm e}}{R_{\rm E}}\right)^2.
\end{equation}
Here $\rm V_e$ is an effective source velocity, def\mbox{}ined as the change in time of the source position measured by the observer, and $R_{\rm E}$ is the Einstein radius,
\begin{equation}
R_{\rm E}=\left( \frac{4GM}{c^2}\frac{D_{\rm{LS}}D_{\rm S}}{D_{\rm L}}\right )^{1/2}, 
\end{equation}
where $D_{\rm{LS}}$, $D_{\rm S}$ and $D_{\rm L}$ are the lens--source, the source--observer and
the lens--observer angular diameter distances, respectively.
\footnote{We assume a flat universe with $\Omega_{\rm m}=0.3$, $\Omega_{\Lambda}=0.7$ and Hubble constant $H_0=70$ $\rm km\ s^{-1}\ Mpc^{-1}$}

In Table \ref{table:2} we show the values of possible microlens  masses and the corresponding transverse velocities are given by equations (1) and (2).
In these calculations we assume that the microlensing is caused by a single 
compact object. Even with that simplifying assumption,
  the lens mass can not be determined without knowing the transverse velocity.

There are two possible limits. Due to short duration of the events the microlenses may actually be nanolenses (planets) 
with masses of order of $10^{-3}M_{\odot}$ for typical transverse velocities of the source of 200--700 km~s$^{-1}$. 
Similar results were obtained for QSO 0957+561 \citep{pelt98}. Conversely,
fast microlensing variability can be caused by a solar mass object magnifying small 
continuum parts of the quasar with relativistic velocities (hot spots) \citep{Sch3} or smooth accretion disk 
occulted by optically-thick, fast moving clouds \citep{Stuart}.  Applying this to the two quasars, for microlens 
mass 1$M_{\odot}$, we obtain source (hot spot, cloud) transverse velocities of $0.08c$  for SBS~1520+530 and 
$0.23c$ for FBQ~0951+2635.

Because of the mass-velocity degeneracy none of these possibilities can be eliminated from a simple analysis. Moreover, we stress that 
mass determination using equation (1) is only possible for single isolated  microlensing events and when the transverse velocity is known.
More detailed analyses \citep{Lew,WY01} assume that the quasar light is affected by numerous compact objects in 
the foreground galaxy, creating a complex network of caustic lines, and not by a single object \citep{Pac86}.

\begin{table}[!t]

\caption{Mass-speed relation for SBS 1520+530 and  FBQ 0951+2635}
\label{table:2}
\centering
\begin{tabular}{c c c}
\hline\hline

Mass & SBS1520+530 & FBQ 0951+2635 \\
$[M_{\odot}]$&Velocity [km/s]&Velocity [km/s]  \\
\hline
$1.0$& $2.44\times 10^4$& $6.87\times 10^4$ \\ 
$0.001$& $771$& 2171\\
$0.0001$ & $244$ &  $687$ \\
\hline
\end{tabular}
\end{table}

We conclude that, of the systems we have studied, those with detected flux varability due to microlensing
are also reported to have intensity anomalies of spectra continuum and emission line differences. For
these systems we have found small linear trends of $\sim 0.005$--0.010 mag/100 days and bumps with amplitude
of $\sim 0.05$ mag and durations of $\sim 100$ days. Interestingly, the systems for which no microlensing 
variability was detected have no convicing detected microlensing from other methods, although there may 
be indications in e.g.\ HE~2149$-$2745 \citep{BC02}. However, smaller microlensing signals ($\sim 0.005$ mag/100 days)
in these systems would have been undetectable in our observations.

\textit{Acknowledgments}.
We thank Joachim Wambsganss, Rolf Stabell, Cecile Faure  and Jaan Pelt for valuable 
comments.
The Dark Cosmology Centre is funded by the DNRF. This work was carried out within the framework of the EC FP6 Marie Curie Research
Training Network ``Astrophysics Network for Galaxy LEnsing Studies (ANGLES)''. 
DP acknowledges receipt of a research studentship at the Nordic Optical Telescope.
The data used are based on observations made with the Nordic Optical Telescope, operated
on the island of La Palma jointly by Denmark, Finland, Iceland,
Norway, and Sweden, in the Spanish Observatorio del Roque de los
Muchachos of the Instituto de Astrof\mbox{}isica de Canarias.

\newpage
\appendix
\scriptsize

\begin{center}
\begin{table*}
{\bf Table.3.} Photometry of  two images of FBQ 0951+2635 quasar and two reference stars.
\centering
\begin{tabular}{|c|c|c|c|c|c|c|c|c|}
\hline
$imag_A$&$imag_B$&$Err Rmag_A$&$Err Imag_B$&$Rmag_{ref1}$&$Err Rmag_{ref1}$&$Rmag_{ref2}$&$Err Rmag_{ref2}$&JD\\
\hline
16.992709 & 18.034618 & 0.0122744& 0.0126427 &19.120347 &0.013807705 &19.482423 &0.014300140 &2451176.75\\
16.949270 & 17.978877 & 0.0116044& 0.0119707 &19.113149 &0.013831616 &19.492741 &0.015953711 &2451243.56\\
16.920836 & 17.940096 & 0.0078402& 0.0085687 &19.130056 &0.013900347 &19.478551 &0.016928954 &2451284.35\\
16.921943 & 17.921308 & 0.0106942& 0.0111525 &19.124617 &0.013865446 &19.488171 &0.016369753 &2451293.41\\
16.903616 & 17.904475 & 0.0117503& 0.0122057 &19.124680 &0.013849218 &19.524657 &0.015659789 &2451312.44\\
16.862798 & 17.944382 & 0.0104246& 0.0109116 &19.134758 &0.013842289 &19.476014 &0.015616346 &2451457.71\\
16.867615 & 17.936585 & 0.0103437& 0.0108806 &19.094160 &0.014071862 &19.418735 &0.022082521 &2451485.68\\
16.880539 & 17.931990 & 0.0039793& 0.0051447 &19.106263 &0.013856345 &19.421328 &0.015121674 &2451538.77\\
16.865312 & 17.907113 & 0.0123886& 0.0127311 &19.147895 &0.013857504 &19.470908 &0.016453767 &2451547.63\\
16.855441 & 17.873516 & 0.0125382& 0.0127244 &19.105739 &0.013940485 &19.501384 &0.022208684 &2451587.46\\
16.841846 & 17.895178 & 0.0129032& 0.0130755 &19.119795 &0.013814855 &19.504479 &0.015086975 &2451610.56\\
16.844951 & 17.885690 & 0.0080746& 0.0085439 &19.124043 &0.013794309 &19.468493 &0.014835601 &2451623.40\\
16.861660 & 17.898676 & 0.0105123& 0.0110129 &19.104203 &0.013865219 &19.456326 &0.015949634 &2451675.41\\
16.863279 & 17.899549 & 0.0103627& 0.0106677 &19.124836 &0.013788088 &19.499720 &0.014183901 &2451682.38\\
16.880199 & 17.916674 & 0.0120163& 0.0123721 &19.112471 &0.013880340 &19.519183 &0.018577257 &2451696.40\\
16.871481 & 17.934753 & 0.0095932& 0.0101394 &19.133569 &0.013859803 &19.490971 &0.016356478 &2451723.39\\
16.891041 & 17.986269 & 0.0109143& 0.0114012 &19.119965 &0.013834873 &19.481175 &0.015792936 &2451821.75\\
16.896003 & 17.955737 & 0.0111720& 0.0113642 &19.136559 &0.013828723 &19.480894 &0.015218081 &2451834.73\\
16.879795 & 17.970869 & 0.0126409& 0.0128128 &19.096878 &0.013967582 &19.436984 &0.017929207 &2451845.73\\
16.894209 & 17.943196 & 0.0042684& 0.0056762 &19.140959 &0.013870139 &19.497979 &0.015667505 &2451851.76\\
16.872078 & 17.953676 & 0.0085913& 0.0091732 &19.126812 &0.013863490 &19.521216 &0.016408040 &2451865.73\\
16.882127 & 17.955942 & 0.0125129& 0.0128133 &19.153717 &0.013810298 &19.505930 &0.015129165 &2451870.76\\
16.888764 & 17.929586 & 0.0122424& 0.0126007 &19.131944 &0.013797467 &19.487918 &0.014854928 &2451880.74\\
16.893100 & 17.945269 & 0.0123475& 0.0126993 &19.114313 &0.013926707 &19.511076 &0.020537456 &2451907.76\\
16.887662 & 17.972687 & 0.0122124& 0.0123627 &19.105902 &0.013820950 &19.475602 &0.014776934 &2451914.55\\
16.896969 & 17.969818 & 0.0095270& 0.0100416 &19.146569 &0.013860617 &19.493764 &0.016385478 &2451923.66\\
16.897377 & 17.958831 & 0.0120538& 0.0124478 &19.135314 &0.013811676 &19.493437 &0.014292094 &2451928.57\\
16.898735 & 17.955738 & 0.0126376& 0.0129884 &19.124129 &0.013826723 &19.503457 &0.014864473 &2451959.52\\
16.884034 & 17.969461 & 0.0042291& 0.0055371 &19.120140 &0.013817373 &19.520684 &0.014550879 &2451967.62\\
16.884285 & 17.954437 & 0.0126648& 0.0130048 &19.144934 &0.013794118 &19.534463 &0.015241120 &2451983.47\\
16.889696 & 17.956140 & 0.0130337& 0.0131508 &19.119829 &0.013884220 &19.533765 &0.017750150 &2451999.36\\
16.886299 & 17.960918 & 0.0134442& 0.0135783 &19.121428 &0.013821902 &19.496912 &0.015105136 &2452000.35\\
16.895846 & 17.932691 & 0.0103642& 0.0106085 &19.094316 &0.013906281 &19.449698 &0.016586867 &2452007.48\\
16.885167 & 17.948287 & 0.0129423& 0.0130618 &19.130773 &0.013812789 &19.525274 &0.014868060 &2452014.38\\
16.898816 & 17.953915 & 0.0086892& 0.0091713 &19.148099 &0.013932938 &19.527538 &0.021234268 &2452016.40\\
16.917229 & 17.951211 & 0.0134910& 0.0136049 &19.147977 &0.013812686 &19.508021 &0.014486790 &2452022.39\\
16.923007 & 17.965599 & 0.0129227& 0.0130598 &19.148200 &0.013777790 &19.500059 &0.014440096 &2452026.42\\
16.906482 & 17.981935 & 0.0111480& 0.0113552 &19.144386 &0.013771457 &19.489314 &0.013947707 &2452030.44\\
16.895825 & 17.965255 & 0.0121185& 0.0123055 &19.085406 &0.013906838 &19.462510 &0.017683878 &2452008.36\\
16.909523 & 17.965818 & 0.0133956& 0.0134637 &19.107931 &0.013836224 &19.481380 &0.016996284 &2452013.42\\
16.922141 & 17.975225 & 0.0103913& 0.0106341 &19.135155 &0.013781629 &19.496244 &0.014553395 &2452033.42\\
16.929039 & 17.991203 & 0.0129533& 0.0130885 &19.127224 &0.013838793 &19.486634 &0.014117430 &2452047.40\\
16.927728 & 18.001496 & 0.0130892& 0.0132157 &19.128467 &0.013777552 &19.514531 &0.014502249 &2452051.39\\
16.929106 & 18.025580 & 0.0131267& 0.0132529 &19.134794 &0.013780329 &19.509152 &0.014461936 &2452055.40\\
16.936453 & 18.009050 & 0.0106027& 0.0108315 &19.159283 &0.013819508 &19.512907 &0.016321199 &2452058.38\\
16.943759 & 17.959676 & 0.0081699& 0.0086049 &19.156133 &0.013809766 &19.522428 &0.015438366 &2452059.38\\
16.933074 & 18.014459 & 0.0068614& 0.0073612 &19.141548 &0.013765144 &19.505601 &0.014014124 &2452060.38\\
16.937465 & 18.054031 & 0.0121674& 0.0123452 &19.141910 &0.013834839 &19.511048 &0.015714603 &2452077.39\\
16.944463 & 18.026573 & 0.0115327& 0.0117283 &19.108928 &0.013834722 &19.455369 &0.015313457 &2452081.38\\
16.851988 & 17.883746 & 0.0062832& 0.0084698 &19.126692 &0.013782970 &19.501844 &0.014435887 &2451466.77\\
16.870338 & 17.913338 & 0.0105393& 0.0111023 &19.126316 &0.013781707 &19.512694 &0.014083847 &2451471.72\\
16.844608 & 17.863082 & 0.0067235& 0.0072235 &19.134932 &0.013783441 &19.511825 &0.014415147 &2451596.57\\
16.866450 & 17.906396 & 0.0045824& 0.0061145 &19.133440 &0.013831234 &19.486610 &0.016333501 &2451703.40\\
16.866219 & 17.944543 & 0.0109743& 0.0112793 &19.101035 &0.013877617 &19.468187 &0.017442770 &2451861.73\\
16.867510 & 17.968184 & 0.0076510& 0.0094711 &19.136351 &0.013891461 &19.503001 &0.017352248 &2451945.68\\
16.880295 & 17.929751 & 0.0040531& 0.0054936 &19.099264 &0.014002759 &19.455817 &0.017569236 &2452003.37\\
16.907873 & 17.972477 & 0.0111268& 0.0114910 &19.137579 &0.013797823 &19.509583 &0.014943108 &2452037.42\\
16.921842 & 17.998561 & 0.0044078& 0.0071272 &19.147866 &0.013811678 &19.487119 &0.015711978 &2452071.38\\
\hline
\end{tabular}
\end{table*}
\end{center}
                                                                                            
\scriptsize
\begin{center}

\begin{table*}  

\centering                                                               
{\bf Table.4.} Photometry of two images of  B1600+434 quasar.

\centering   
                                                                                    
\begin{tabular}{|c|c|c|c|c|}

\hline                                                                                            
$imag_A$&$Err Imag_A$&$imag_B$&$Err Imag_B$&JD\\                                                                    
\hline
5.2819 & 0.0296 & 5.8385 & 0.0356 & 2450906.00\\                                                        
5.3158 & 0.0315 & 6.0376 & 0.0362 & 2450925.00\\                                                        
5.2249 & 0.0324 & 6.2962 & 0.0395 & 2450935.00\\                                                        
5.2928 & 0.0288 & 6.0041 & 0.0402 & 2450944.00\\                                                        
5.2868 & 0.0295 & 6.0137 & 0.0387 & 2450952.00\\                                                        
5.3941 & 0.0295 & 5.9756 & 0.0360 & 2450954.00\\                                                        
5.1570 & 0.0315 & 5.8081 & 0.0356 & 2450957.00\\                                                        
5.3272 & 0.0285 & 5.7904 & 0.0358 & 2450963.00\\                                                        
5.2983 & 0.0283 & 6.0343 & 0.0446 & 2450987.00\\                                                        
5.3121 & 0.0298 & 5.9627 & 0.0367 & 2450994.00\\                                                        
5.3363 & 0.0287 & 6.0556 & 0.0404 & 2451014.00\\                                                        
5.2926 & 0.0291 & 5.9526 & 0.0385 & 2451021.00\\                                                        
5.2632 & 0.0289 & 5.9821 & 0.0398 & 2451036.00\\                                                        
5.1212 & 0.0305 & 5.9744 & 0.0390 & 2451050.00\\                                                        
4.9277 & 0.0301 & 6.1430 & 0.0486 & 2451055.00\\
5.4757 & 0.0284 & 6.0166 & 0.0378 & 2451071.00\\
5.3390 & 0.0283 & 5.7932 & 0.0353 & 2451085.00\\
5.5449 & 0.0292 & 5.8710 & 0.0388 & 2451092.00\\
5.6352 & 0.0420 & 5.7032 & 0.0471 & 2451101.00\\
5.6096 & 0.0461 & 5.6322 & 0.0468 & 2451105.00\\
5.6159 & 0.0329 & 5.6729 & 0.0363 & 2451109.00\\
5.7076 & 0.0604 & 6.2264 & 0.1002 & 2451122.00\\
5.8021 & 0.0805 & 5.9997 & 0.0974 & 2451137.00\\
5.3782 & 0.0297 & 5.8271 & 0.0333 & 2451227.00\\
5.2853 & 0.0291 & 5.9171 & 0.0364 & 2451236.00\\
5.0390 & 0.0297 & 5.9352 & 0.0413 & 2451284.00\\
4.6846 & 0.0300 & 5.6948 & 0.0430 & 2451303.75\\
4.7455 & 0.0283 & 5.5117 & 0.0445 & 2451342.00\\
4.5086 & 0.0322 & 5.2692 & 0.0361 & 2451350.00\\
4.9867 & 0.0321 & 5.1490 & 0.0328 & 2451353.00\\
4.4828 & 0.0327 & 4.6215 & 0.0332 & 2451362.00\\
4.7733 & 0.0422 & 4.2523 & 0.0292 & 2451371.50\\
4.3872 & 0.0333 & 4.8200 & 0.0344 & 2451377.50\\
4.9513 & 0.0290 & 5.5594 & 0.0373 & 2451403.50\\
4.9932 & 0.0297 & 5.5444 & 0.0353 & 2451421.50\\
5.1450 & 0.0300 & 5.3622 & 0.0319 & 2451433.50\\
5.1311 & 0.0303 & 5.3218 & 0.0318 & 2451434.25\\
5.2592 & 0.0283 & 5.8469 & 0.0400 & 2451450.25\\
5.2670 & 0.0315 & 5.9853 & 0.0365 & 2451457.25\\
5.2361 & 0.0446 & 5.7947 & 0.0795 & 2451465.25\\
5.4954 & 0.0295 & 6.0011 & 0.0465 & 2451472.25\\
\hline                                        
\end{tabular} 
                                       
\end{table*}
\end{center}

\newpage
\begin{center}
\scriptsize                                       
\begin{table*}                             
{\bf Table.5.}Photometry of two images of HE 2149-2745 quasar in i-band.                   
\centering                                    
\begin{tabular}{|c|c|c|c|c|}

\hline                                  
$imag_A$&$Err Imag_A$&$imag_B$&$Err Imag_B$&JD\\          
\hline                                        
1.58210003 & 0.0378655307 & 3.06310010 & 0.0389079675 & 2451091.5\\
1.57809997 & 0.0378287919 & 3.07520008 & 0.0383489877 & 2451058.7\\
1.56980002 & 0.0378377251 & 3.04200006 & 0.0384890176 & 2451100.5\\
1.64760005 & 0.0378425010 & 3.10999990 & 0.0385378115 & 2451107.5\\
1.53980005 & 0.0378421955 & 3.01279998 & 0.0385516994 & 2451114.5\\
1.60200000 & 0.0378793776 & 3.07660007 & 0.0391337611 & 2451119.5\\
1.60819995 & 0.0378611907 & 3.06750011 & 0.0388117172 & 2451137.5\\
1.61109996 & 0.0378544070 & 3.05719995 & 0.0387288556 & 2451124.5\\
1.57679999 & 0.0378506258 & 3.03020000 & 0.0386729501 & 2451130.5\\
1.60739994 & 0.0378769375 & 3.08220005 & 0.0391000696 & 2451146.5\\
1.62779999 & 0.0378769971 & 3.09910011 & 0.0390986577 & 2451160.5\\
1.62199998 & 0.0378788561 & 3.06690001 & 0.0390518941 & 2451164.5\\
1.62600005 & 0.0378556550 & 3.04169989 & 0.0387061089 & 2451167.5\\
1.61249995 & 0.0378501825 & 3.06040001 & 0.0386340357 & 2451153.5\\
1.69700003 & 0.0378291346 & 3.13479996 & 0.0383005626 & 2451315.0\\
1.69659996 & 0.0379458442 & 3.13969994 & 0.0400222652 & 2451333.7\\
1.66649997 & 0.0378436223 & 3.12859988 & 0.0385885425 & 2451349.0\\
1.70169997 & 0.0378417931 & 3.14669991 & 0.0384832397 & 2451309.0\\
1.68959999 & 0.0378342941 & 3.13730001 & 0.0383743569 & 2451321.0\\
1.68879998 & 0.0378567949 & 3.14389992 & 0.0387759842 & 2451367.2\\
1.68270004 & 0.0378336124 & 3.12899995 & 0.0383646935 & 2451424.5\\
1.70990002 & 0.0378260836 & 3.15849996 & 0.0382536501 & 2451431.7\\
1.71210003 & 0.0378338359 & 3.17630005 & 0.0383979045 & 2451379.7\\
1.66729999 & 0.0378314964 & 3.16689992 & 0.0383910239 & 2451382.7\\
1.69579995 & 0.0378432684 & 3.18959999 & 0.0385701284 & 2451438.5\\
1.73540003 & 0.0378137380 & 3.26299996 & 0.0392147452 & 2451844.5\\
1.69760001 & 0.0378320850 & 3.15790009 & 0.0383670107 & 2451450.5\\
1.67579997 & 0.0378207192 & 3.22160006 & 0.0382636935 & 2451396.7\\
1.70350003 & 0.0378232971 & 3.17950010 & 0.0382370576 & 2451402.5\\
1.69690001 & 0.0378274396 & 3.18779993 & 0.0383255184 & 2451408.5\\
1.68819997 & 0.0378242321 & 3.20910010 & 0.0382717140 & 2451791.7\\
1.72570002 & 0.0378414802 & 3.19670010 & 0.0385198407 & 2451462.5\\
1.68330002 & 0.0379273593 & 3.15470004 & 0.0398344211 & 2451471.5\\
1.69990003 & 0.0378282033 & 3.21900010 & 0.0383800194 & 2451479.5\\
1.71200001 & 0.0378198288 & 3.20230007 & 0.0381722786 & 2451484.5\\
1.73169994 & 0.0378440991 & 3.21190000 & 0.0385658108 & 2451519.5\\
1.71099997 & 0.0378429405 & 3.23639989 & 0.0386381000 & 2451823.7\\
1.67690003 & 0.0378333218 & 3.19930005 & 0.0384436920 & 2451831.5\\
1.68209997 & 0.0378317647 & 3.18689993 & 0.0384052061 & 2451750.7\\
1.70089996 & 0.0379198268 & 3.22169995 & 0.0398468710 & 2451754.7\\
\hline
\end{tabular}

\end{table*}
\end{center}
\newpage
\scriptsize
\begin{center}
\begin{table*}

\centering
{\bf Table.6.} Photometry of two images of quasar HE 2149-2745 in V-band.

\centering
\begin{tabular}{|c|c|c|c|c|}

\hline
$imag_A$&$Err Imag_A$&$imag_B$&$Err Imag_B$&JD\\
\hline
- 0.0304 & 0.0209 & 1.6168 & 0.0224 & 2451091.50\\
- 0.0372 & 0.0210 & 1.5774 & 0.0213 & 2451058.75\\
- 0.0231 & 0.0210 & 1.5796 & 0.0217 & 2451101.50\\
- 0.0070 & 0.0210 & 1.5987 & 0.0232 & 2451119.50\\
- 0.0131 & 0.0210 & 1.5835 & 0.0216 & 2451124.50\\
- 0.0006 & 0.0210 & 1.6006 & 0.0217 & 2451130.50\\
  0.0017 & 0.0210 & 1.5937 & 0.0216 & 2451137.50\\
- 0.0516 & 0.0210 & 1.5411 & 0.0219 & 2451107.50\\
  0.1087 & 0.0210 & 1.7007 & 0.0214 & 2451396.75\\
  0.0001 & 0.0210 & 1.6068 & 0.0221 & 2451146.50\\
  0.0060 & 0.0210 & 1.6004 & 0.0217 & 2451153.50\\
  0.0218 & 0.0210 & 1.6075 & 0.0217 & 2451160.50\\
  0.0205 & 0.0210 & 1.6017 & 0.0218 & 2451164.50\\
  0.0165 & 0.0210 & 1.5885 & 0.0218 & 2451167.50\\
  0.1169 & 0.0210 & 1.7016 & 0.0221 & 2451300.00\\
  0.1146 & 0.0210 & 1.6862 & 0.0215 & 2451309.00\\
  0.1075 & 0.0210 & 1.6941 & 0.0215 & 2451316.00\\
  0.1027 & 0.0210 & 1.7046 & 0.0215 & 2451321.00\\
  0.0983 & 0.0210 & 1.7812 & 0.0245 & 2451333.75\\
  0.0899 & 0.0210 & 1.7395 & 0.0215 & 2451340.00\\
  0.0989 & 0.0210 & 1.7336 & 0.0218 & 2451366.75\\
  0.0932 & 0.0210 & 1.7315 & 0.0216 & 2451379.75\\
  0.0957 & 0.0210 & 1.7592 & 0.0217 & 2451382.75\\
  0.1015 & 0.0210 & 1.7374 & 0.0215 & 2451402.75\\
  0.1054 & 0.0210 & 1.7468 & 0.0216 & 2451423.75\\
  0.1102 & 0.0210 & 1.7361 & 0.0216 & 2451438.50\\
  0.1142 & 0.0210 & 1.7321 & 0.0215 & 2451431.75\\
  0.0935 & 0.0210 & 1.7491 & 0.0215 & 2451407.50\\
  0.1128 & 0.0210 & 1.7582 & 0.0255 & 2451442.50\\
  0.1088 & 0.0210 & 1.7407 & 0.0218 & 2451450.50\\
  0.1222 & 0.0210 & 1.7605 & 0.0216 & 2451462.25\\
  0.1327 & 0.0210 & 1.7295 & 0.0224 & 2451470.50\\
  0.1172 & 0.0210 & 1.7575 & 0.0216 & 2451477.50\\
  0.1147 & 0.0210 & 1.7500 & 0.0215 & 2451478.50\\
  0.1176 & 0.0210 & 1.6863 & 0.0213 & 2451485.50\\
  0.1187 & 0.0210 & 1.7534 & 0.0212 & 2451493.50\\
  0.1251 & 0.0210 & 1.7632 & 0.0215 & 2451498.50\\
  0.1382 & 0.0210 & 1.7414 & 0.0218 & 2451513.50\\
  0.1350 & 0.0210 & 1.7638 & 0.0230 & 2451518.50\\
  0.1396 & 0.0210 & 1.7496 & 0.0215 & 2451526.50\\
  0.0815 & 0.0210 & 1.7651 & 0.0238 & 2451722.00\\
  0.0830 & 0.0210 & 1.7375 & 0.0212 & 2451726.75\\
  0.0864 & 0.0210 & 1.7666 & 0.0217 & 2451705.75\\
  0.0707 & 0.0210 & 1.7447 & 0.0225 & 2451740.25\\
  0.0838 & 0.0210 & 1.7375 & 0.0212 & 2451726.50\\
  0.0834 & 0.0210 & 1.7584 & 0.0217 & 2451697.75\\
  0.0810 & 0.0210 & 1.7625 & 0.0217 & 2451750.75\\
  0.0728 & 0.0210 & 1.7313 & 0.0216 & 2451754.75\\
  0.0630 & 0.0210 & 1.7376 & 0.0218 & 2451764.75\\
  0.0364 & 0.0210 & 1.6978 & 0.0227 & 2451801.50\\
  0.0416 & 0.0210 & 1.7115 & 0.0217 & 2451810.50\\
  0.0262 & 0.0210 & 1.6997 & 0.0222 & 2451831.50\\
  0.0366 & 0.0210 & 1.6977 & 0.0213 & 2451837.50\\
  0.0302 & 0.0210 & 1.7109 & 0.0213 & 2451844.50\\
  0.0441 & 0.0210 & 1.7261 & 0.0328 & 2451867.50\\
  0.0344 & 0.0210 & 1.6835 & 0.0224 & 2451884.50\\
  0.0353 & 0.0210 & 1.6734 & 0.0217 & 2451876.50\\
\hline
\end{tabular}
\end{table*}
\end{center}
\newpage
\scriptsize
\begin{center}

\begin{table*}

\centering
{\bf Table.7.} Photometry of two images of quasar RX J0911+0551.

\centering
\begin{tabular}{|c|c|c|c|c|}

\hline
$imag_A$&$Err Imag_A$&$imag_B$&$Err Imag_B$&JD\\
\hline
17.4314 & 0.012097 & 19.2829 & 0.012846 & 2450513.75\\
17.1167 & 0.012261 & 18.8402 & 0.013740 & 2450769.75\\
17.1257 & 0.013128 & 18.9104 & 0.018604 & 2450784.75\\
17.0083 & 0.011679 & 18.9664 & 0.015633 & 2450833.50\\
16.9780 & 0.012216 & 18.9863 & 0.014306 & 2450952.00\\
16.9878 & 0.012588 & 18.9918 & 0.017682 & 2450955.50\\
16.9855 & 0.012432 & 18.9854 & 0.016262 & 2450957.50\\
17.0099 & 0.018110 & 19.0251 & 0.049982 & 2450978.50\\
17.0016 & 0.016199 & 19.1349 & 0.044582 & 2450994.50\\
17.0099 & 0.012700 & 19.0251 & 0.018789 & 2451039.50\\
17.0989 & 0.012273 & 19.2174 & 0.015305 & 2451136.75\\
17.1774 & 0.012200 & 19.3156 & 0.014460 & 2451162.75\\
17.2081 & 0.012094 & 19.2239 & 0.012973 & 2451177.75\\
17.2514 & 0.012870 & 19.2023 & 0.019072 & 2451235.00\\
17.2558 & 0.012188 & 19.1749 & 0.013675 & 2451243.00\\
17.2838 & 0.012242 & 19.1644 & 0.013936 & 2451256.00\\
17.3230 & 0.012786 & 19.0844 & 0.016953 & 2451283.00\\
17.3340 & 0.012643 & 19.0819 & 0.016079 & 2451293.00\\
17.3293 & 0.012319 & 19.0868 & 0.014112 & 2451306.50\\
17.3090 & 0.013496 & 19.0787 & 0.020718 & 2451312.50\\
17.2628 & 0.014121 & 19.0474 & 0.024044 & 2451321.50\\
17.2391 & 0.012372 & 19.0314 & 0.014646 & 2451330.00\\
17.2386 & 0.012744 & 19.0592 & 0.017307 & 2451336.50\\
17.2382 & 0.012927 & 19.0693 & 0.018480 & 2451342.50\\
17.2060 & 0.013608 & 19.0799 & 0.022970 & 2451353.50\\
17.2071 & 0.013444 & 19.0877 & 0.022186 & 2451355.50\\
17.0963 & 0.012342 & 19.1293 & 0.015669 & 2451456.75\\
17.0719 & 0.012927 & 19.1268 & 0.020853 & 2451465.75\\
17.0660 & 0.012393 & 19.1474 & 0.016441 & 2451470.75\\
17.0987 & 0.012193 & 19.2220 & 0.014468 & 2451484.75\\
17.1064 & 0.012217 & 19.2785 & 0.015011 & 2451511.75\\
17.1368 & 0.012188 & 19.2781 & 0.014460 & 2451530.75\\
17.1294 & 0.012145 & 19.2944 & 0.013966 & 2451546.75\\
17.1312 & 0.014747 & 19.2710 & 0.035811 & 2451566.75\\
17.1260 & 0.012167 & 19.3209 & 0.014371 & 2451587.50\\
17.1453 & 0.012479 & 19.2673 & 0.017630 & 2451595.50\\
17.1542 & 0.012282 & 19.2853 & 0.015319 & 2451610.50\\
17.1049 & 0.012669 & 19.3130 & 0.020335 & 2451613.50\\
17.0979 & 0.012658 & 19.2729 & 0.019833 & 2451613.50\\
17.1058 & 0.012671 & 19.3169 & 0.020383 & 2451613.50\\
17.1481 & 0.012445 & 19.2519 & 0.017215 & 2451623.50\\
17.2804 & 0.012584 & 19.2286 & 0.017100 & 2451675.50\\
17.2842 & 0.012405 & 19.2229 & 0.015676 & 2451682.25\\
17.2756 & 0.013434 & 19.2272 & 0.023092 & 2451690.25\\
17.2754 & 0.012512 & 19.2225 & 0.016512 & 2451696.50\\
17.2040 & 0.012910 & 19.3341 & 0.021965 & 2451798.75\\
17.1908 & 0.012287 & 19.3688 & 0.015887 & 2451814.75\\
17.1812 & 0.012463 & 19.3403 & 0.017784 & 2451820.75\\
17.1682 & 0.012275 & 19.2998 & 0.015553 & 2451833.75\\
17.1901 & 0.012176 & 19.3068 & 0.014179 & 2451843.75\\
17.1850 & 0.012226 & 19.2968 & 0.014778 & 2451844.75\\
17.2105 & 0.012952 & 19.3215 & 0.021781 & 2451850.75\\
\hline
\multicolumn{5}{|c|}{continued on next page.}\\
\hline
\end{tabular}                               
\end{table*}
\end{center} 
\begin{center}

\begin{table*}

\centering
\begin{tabular}{|c|c|c|c|c|}
\hline
\multicolumn{5}{|c|}{continued from previous page.}\\
\hline
$imag_A$&$Err Imag_A$&$imag_B$&$Err Imag_B$&JD\\
\hline
17.2203 & 0.012489 & 19.1839 & 0.016541 & 2451864.75\\
17.2371 & 0.012213 & 19.2037 & 0.013989 & 2451869.75\\
17.2566 & 0.012470 & 19.1875 & 0.016057 & 2451879.75\\
17.2626 & 0.012195 & 19.1742 & 0.013649 & 2451906.75\\
17.2484 & 0.012238 & 19.1593 & 0.014067 & 2451913.50\\
17.2460 & 0.012263 & 19.1749 & 0.014233 & 2451922.50\\
17.2528 & 0.012202 & 19.1022 & 0.013505 & 2451927.50\\
17.2591 & 0.012341 & 19.1918 & 0.014997 & 2451935.50\\
17.2820 & 0.012366 & 19.1509 & 0.014891 & 2451951.50\\
17.2727 & 0.012207 & 19.1396 & 0.013631 & 2451958.50\\
17.2574 & 0.012254 & 19.1466 & 0.014101 & 2451966.50\\
17.2570 & 0.012309 & 19.1568 & 0.014555 & 2451966.50\\
17.2437 & 0.012180 & 19.1896 & 0.013604 & 2451983.50\\
17.2155 & 0.029405 & 19.2627 & 0.074262 & 2451990.25\\
17.2330 & 0.012209 & 19.2268 & 0.014023 & 2451998.50\\
17.2372 & 0.012308 & 19.2229 & 0.014987 & 2451999.25\\
17.2247 & 0.012686 & 19.2174 & 0.017995 & 2451999.75\\
17.2250 & 0.012686 & 19.1970 & 0.017995 & 2452000.25\\
17.2093 & 0.013035 & 19.2109 & 0.020981 & 2452006.50\\
17.1904 & 0.012433 & 19.2521 & 0.016423 & 2452014.50\\
17.1750 & 0.012724 & 19.2797 & 0.019634 & 2452016.50\\
17.1454 & 0.006381 & 19.2605 & 0.043823 & 2452021.50\\
\hline          
\end{tabular}                               
\end{table*}
\end{center}

\newpage
\begin{center}                                                
\begin{table*}                               
{\bf Table.8.} Photometry of two images of SBS 1520+530 quasar.                        
\centering                                      

\begin{tabular}{|c|c|c|c|c|}

\hline                                    
$imag_A$&$Err Imag_A$&$imag_B$&$Err Imag_B$&JD\\    
\hline

1.2548 & 0.01430 & 1.8684 & 0.02422 & 2451226.75\\      
1.1809 & 0.00911 & 1.8650 & 0.01154 & 2451235.75\\      
1.2221 & 0.00876 & 1.8486 & 0.00967 & 2451243.75\\      
1.2245 & 0.00875 & 1.8931 & 0.00970 & 2451255.75\\      
1.1905 & 0.00869 & 1.8935 & 0.00966 & 2451264.75\\      
1.2312 & 0.00869 & 1.9357 & 0.00957 & 2451284.50\\      
1.2257 & 0.00873 & 1.9430 & 0.00981 & 2451306.50\\      
1.2154 & 0.00879 & 1.9373 & 0.01016 & 2451311.50\\      
1.1879 & 0.00871 & 1.9115 & 0.00978 & 2451319.50\\      
1.1747 & 0.00867 & 1.9314 & 0.00960 & 2451336.50\\      
1.1725 & 0.00873 & 1.9221 & 0.00997 & 2451342.50\\      
1.1629 & 0.00871 & 1.9154 & 0.00978 & 2451354.50\\      
1.1426 & 0.00859 & 1.9055 & 0.00911 & 2451362.50\\      
1.1448 & 0.00891 & 1.9151 & 0.01090 & 2451371.50\\      
1.1341 & 0.00860 & 1.9162 & 0.00922 & 2451377.50\\      
1.1426 & 0.00867 & 1.9214 & 0.00960 & 2451386.50\\      
1.1525 & 0.00877 & 1.9316 & 0.01021 & 2451403.50\\      
1.1515 & 0.00864 & 1.9627 & 0.00953 & 2451420.25\\      
1.1293 & 0.00857 & 1.9227 & 0.00900 & 2451433.25\\      
1.1310 & 0.00856 & 1.9197 & 0.00898 & 2451434.25\\      
1.1341 & 0.00898 & 1.8800 & 0.00962 & 2451443.25\\      
1.1406 & 0.00860 & 1.8994 & 0.00916 & 2451450.25\\      
1.1449 & 0.00877 & 1.8658 & 0.00999 & 2451457.25\\      
1.1433 & 0.00892 & 1.8503 & 0.01074 & 2451465.25\\      
1.0952 & 0.00871 & 1.8334 & 0.00966 & 2451530.75\\      
1.1086 & 0.00867 & 1.8229 & 0.00944 & 2451566.75\\      
1.1139 & 0.00870 & 1.8227 & 0.00963 & 2451586.75\\      
1.0913 & 0.00868 & 1.8227 & 0.00959 & 2451609.50\\
1.1137 & 0.00890 & 1.8154 & 0.01048 & 2451613.75\\
1.1094 & 0.00869 & 1.7793 & 0.00961 & 2451674.50\\
1.1107 & 0.00869 & 1.7898 & 0.00947 & 2451683.50\\
1.1174 & 0.00883 & 1.7758 & 0.01016 & 2451689.50\\
1.1111 & 0.00867 & 1.8072 & 0.00953 & 2451695.50\\
1.0908 & 0.00879 & 1.7812 & 0.01000 & 2451703.50\\
1.0908 & 0.01052 & 1.8043 & 0.01639 & 2451715.50\\
1.0869 & 0.00866 & 1.7966 & 0.00955 & 2451723.50\\
1.0907 & 0.00866 & 1.8130 & 0.00961 & 2451730.50\\
1.1080 & 0.00861 & 1.8403 & 0.00918 & 2451751.50\\
1.0948 & 0.00871 & 1.8317 & 0.00974 & 2451765.50\\
1.0880 & 0.00859 & 1.8285 & 0.00913 & 2451781.50\\
1.0747 & 0.00860 & 1.8415 & 0.00913 & 2451793.50\\
1.0588 & 0.00857 & 1.8399 & 0.00900 & 2451799.25\\
1.0616 & 0.00858 & 1.8407 & 0.00902 & 2451807.50\\
1.0468 & 0.00877 & 1.8561 & 0.01026 & 2451815.25\\
1.0013 & 0.00864 & 1.8408 & 0.00948 & 2451834.25\\
0.9944 & 0.00882 & 1.8408 & 0.01070 & 2451869.75\\
1.1008 & 0.00867 & 1.8286 & 0.00945 & 2451907.75\\
1.1141 & 0.00857 & 1.8240 & 0.00891 & 2451913.75\\
1.1446 & 0.00863 & 1.8035 & 0.00911 & 2451922.75\\
1.1587 & 0.00871 & 1.7976 & 0.00950 & 2451927.75\\
1.1866 & 0.00865 & 1.7408 & 0.00904 & 2451950.75\\
1.1964 & 0.00871 & 1.7304 & 0.00934 & 2451958.75\\
1.2070 & 0.00876 & 1.7117 & 0.00938 & 2451966.50\\
1.2178 & 0.00899 & 1.7128 & 0.00995 & 2451975.75\\
1.2139 & 0.00864 & 1.7222 & 0.00896 & 2451983.75\\
1.2220 & 0.00865 & 1.7232 & 0.00898 & 2451984.75\\
1.2168 & 0.00901 & 1.7594 & 0.01011 & 2452007.50\\
1.2446 & 0.00874 & 1.7632 & 0.00932 & 2452007.75\\
\hline                            
\end{tabular}                      
\end{table*}
\end{center}                        
\end{document}